\documentclass[showpacs,aps,pra,floatfix,reprint,superscriptaddress,footinbib,citeautoscript]{revtex4-2}
\usepackage{amssymb,amsfonts,amsmath}
\usepackage{graphicx}
\usepackage{verbatim}
\usepackage[utf8]{inputenc}
\usepackage[T1]{fontenc}
\usepackage{soul,xcolor,colortbl}
\usepackage{multirow}
\usepackage{bm}
\usepackage{array}
\usepackage{chemmacros}
\usepackage{color}
\usepackage{hyperref} \hypersetup{colorlinks=true,citecolor=blue,linkcolor=blue,urlcolor=blue}
\usepackage{mdframed}
\usepackage{longtable}
\usepackage{enumitem}
\usepackage{booktabs}
\usepackage{lipsum}
\usepackage{orcidlink}

\newcolumntype{L}[1]{>{\raggedright\arraybackslash}p{#1} }
\newcolumntype{C}[1]{>{\centering \arraybackslash}p{#1} }
\newcolumntype{R}[1]{>{\raggedleft \arraybackslash}p{#1} }

\DeclareSIUnit\angstrom{\text{\AA}}

\newlist{noteize}{itemize}{4}
\setlist[noteize]{label={\textbullet}, font=\footnotesize, noitemsep, align=parleft, labelwidth=0.5em, leftmargin=1em}

\def\AFLOW{{\small AFLOW}}

\def\DFT{{\small DFT}}
\def\DVC{{\small DVC}}
\def\DFTDVC{{\small DFT+DVC}}

\def\POCC{{\small POCC}}
\def\VASP{{\small VASP}}
\def\PBE{{\small PBE}}
\def\ECI{{\small ECI}}
\def\ECIs{{\small ECI}s}

\def\sPOCC{{\substack{\scalebox{0.5}{POCC}}}}
\def\ssPOCC{{\substack{\scalebox{0.4}{POCC}}}}
\def\ssp{\mathrm{sp}}
\def\sgrid{{\substack{\scalebox{0.5}{grid}}}}

\def\smix{\mathrm{mix}}
\def\ssi{\mathrm{si}}
\def\Essi{\mathcal{E}_\ssi}
\def\CALPHAD{{\small CALPHAD}}

\def\XXX{{Methods}}

\setcitestyle{square}

\makeatletter \renewcommand\frontmatter@abstractwidth{\dimexpr\textwidth\relax} \makeatother

\def\MEMS{\footnotesize Department of Mechanical Engineering and Materials Science, Duke University, Durham, NC 27708, USA}
\def\CEM{\footnotesize Center for Extreme Materials, Duke University, Durham, NC 27708, USA}

\begin{document}

\title{Disorder viscosity correction approach to calculate spinodal temperature and wavelength}

\author{Simon~Divilov\,\orcidlink{0000-0002-4185-6150}}\affiliation{\MEMS}\affiliation{\CEM}
\author{Hagen~Eckert\,\orcidlink{0000-0003-4771-1435}}\affiliation{\MEMS}\affiliation{\CEM}
\author{Nico Hotz\,\orcidlink{0009-0008-2469-2693}}\affiliation{\MEMS}\affiliation{\CEM}
\author{Xiomara~Campilongo\,\orcidlink{0000-0001-6123-8117}}\affiliation{\CEM}
\author{Stefano~Curtarolo\,\orcidlink{0000-0003-0570-8238}}\email[]{stefano@duke.edu}\affiliation{\MEMS}\affiliation{\CEM}

\date{\today}

\begin{abstract}\noindent
Spinodal decomposition, a key mechanism to microstructure formation in materials, has long posed challenges for predictive modeling, due to the need for parameter-free approaches that accurately capture local energy landscapes.
In this work, we propose an approach to predict spinodal behavior by introducing a disorder viscosity correction to bulk free energies computed from finite, small, representative cells.
We approximate the energy penalty required to transition into a disordered state to enable the stabilization of locally concave bulk free energy regions --- essential for interface formation --- while suppressing long-range concentration fluctuations.
This approximation circumvents the complexity of full ab initio parameterization of interfacial properties and is well-suited for high-throughput and machine-learning frameworks.
Our approach captures the necessary physics underpinning spinodal kinetics, offering a scalable route to predict spinodal regions in compositionally complex and high-entropy materials.
\end{abstract}
\maketitle

\noindent

\section{Introduction}
Spinodal decomposition is a type of phase segregation characterized by a periodic composition modulation occuring after quenching a solid solution within the spinodal region, a sub-region of the miscibility gap~\cite{Cahn1961795}.
The presence of this microstructure can improve many properties~\cite{curtarolo:art80_etal} like hardness, magnetic coercivity, magnetoresistance, thermoelectric performance, plasticity, and ductility~\cite{Cahn_ActaMetall_SpinodalHardening_1963,Findik_MATTDES_2012,Dietl_RMP_2015_etal}.
Unfortunately, predicting the onset of spinodal decomposition without any experimentally-inherited parameters (i.e., phase-field~\cite{Marro_PRB_1975,Langer_PRA_1975}{, \CALPHAD~\cite{Xiong_SSP_2011,Zhou_CALPHAD_2017} or machine-learning~\cite{Deffrennes_MATTDES_2022}}) has been notoriously difficult.
{
On the other hand, parameter-free procedures such as interatomic potentials~\cite{curtarolo:art201_etal}, cluster expansion~\cite{deFontaine_ssp_1994}, and the cluster variation method~\cite{kikuchi} are not easily applicable to high-throughput workflows~\cite{nmatHT,curtarolo:art114_etal}, which require a large amount of computational resources -- especially for multi-component disordered systems~\cite{clupan}.
}

This difficulty~\cite{deFontaine_Clustering_1975} stems from the fact that existing theories~\cite{kikuchi,Korringa1947392} attempt to model the disordered system as a state in thermodynamic equilibrium, where the bulk free energy, as a function of concentration, $F(\mathbf{x})$ is globally convex, without multiple minima~\cite{Landau_CTP5_SP_1969} --- i.e., no spinodal decomposition can exist.
However, real materials contain defects, grains, interfaces or sluggish kinetics which effectively impede large and long-range compositional fluctuations, hindering the reach of the ideal equilibrium~\cite{Kinetic_of_Materials}.
As such, while the \textit{global} $F$ remains a convex function, the interfaces and other distortive effects can stabilize a local concave \textit{bulk} $F$ --- having metastable minima --- inducing a multitude of phenomena, among them, spinodal kinetics~\cite{Langer_AnnPhys_1971,cahn_hilliard_1}.
\begin{figure*}
    \includegraphics[width=\linewidth]{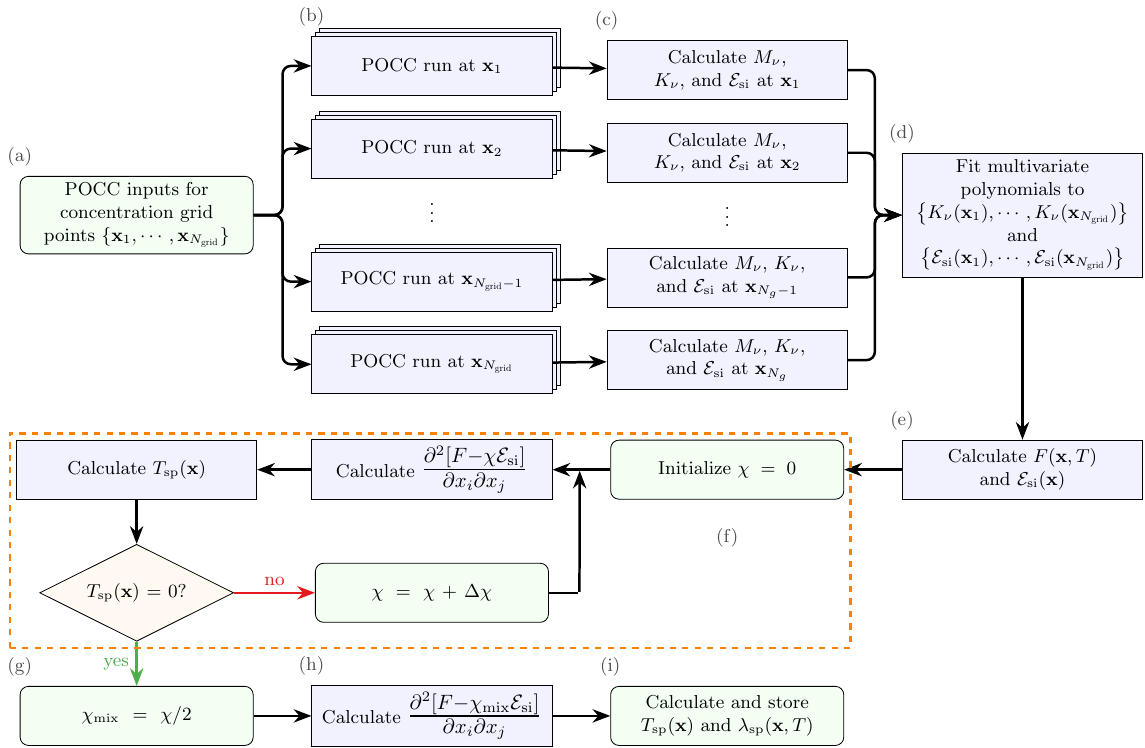}
    \caption{\small
    \textbf{Disorder viscosity correction workflow for calculating spinodal temperature and wavelength.} The workflow describes the approach to calculate the spinodal temperature and wavelength using the \underline{d}isorder \underline{v}iscosity \underline{c}orrection (\DVC).
    {\bf (a)}
    We begin by generating \POCC\ tiles of the minimal size congruent with a concentration grid $\{\mathbf{x}_1,\cdots,\mathbf{x}_{N_\sgrid}\}$ of $N_\sgrid$ points and calculating their ab initio energies.
    {\bf (b)}
    Then, using the \AFLOW\ standard (\XXX), we calculate the \DFT\ energies of all the \POCC\ tiles.
    {\bf (c)}
    Next, at each concentration grid point, the moments $M_\nu$, cumulants $K_\nu$ and self-interacting energy of the disordered system $\Essi$ are computed.
    The self-interacting energy is extremely difficult to calculate, and therefore, in the zeroth approximation and infinite-temperature limit, we approximate it by the ensemble average energy of the disordered system --- e.g., $\Essi\approx\sum_iP_iE_i$, where $P_i$ and $E_i$ are the probability and energy of the $i$-tile, respectively.
    {\bf (d)}
    Afterwards, we fit multivariate polynomials in $\mathbf{x}$ for $K_\nu$ and $\Essi$.
    {\bf (e)}
    This allows us to obtain analytical expressions for $F$ and $\Essi$.
    As discussed in the text, the choice in the size of \POCC\ tiles leads to an over-stabilization of the spinodal temperature $T_\ssp$.
    To remedy this issue, we correct $F$ by a fraction of the self-interaction energy $\chi\Essi$, where $\chi\in [0,1]$.
    We call this the disorder viscosity correction.
    {\bf (f)}
    The \DVC\ is done self-consistently through the calculation of $\chi_\smix$ (orange dashed box).
    {\bf (g)}
    We determine $\chi_\smix$ as the average (1/2) between no correction ($\chi\equiv0$) and the value at which the system is completely miscible (the minimum $\chi$ value at which $T_\ssp$ is zero for the whole concentration space).
    {\bf (h)}
    With $\chi_\smix$, we calculate the concentration-Hessian of the corrected free energy $F-\chi_\smix\Essi$.
    {\bf (i)}
    Finally, we use this to compute the corrected $T_\ssp$ and spinodal wavelength $\lambda_\ssp$.
    \label{fig:workflow}
    }
\end{figure*}

Building a model capable of capturing the formation of a microstructure having non-infinite dimensions (wavelengths) is not trivial, it requires the description of the interaction between non-flat regions with different concentrations, so that the curvature-stabilizing effect can arise. In an ideal scenario, the free energy of a disordered system phase separates upon cooling~\cite{lupis} and restores its global convexity through long-range compositional fluctuations~\cite{Kikuchi_JCP_1972,Cenedese_PHYSA_1994}, leading to a global reduction of interfaces between coexisting phases~\cite{Kikuchi_JCP_1967}. This is not what happens inside the spinodal region. There, the energetic balance between the lattice distortions and the curvature of the interfaces allows the free energy to recover the global convexity from its local bulk concavity by creating as many curved interfaces as allowed~\cite{BaluffiAllenCarter_book}.

From an ab initio standpoint, this is a daunting task~\cite{deFontaine_Clustering_1975,Cenedese_PHYSA_1994}: it would require the parameterization of surface energies and tensions of enough interface shapes and concentration to get the correct differential behavior of the free energy in the diffusion equation~\cite{BaluffiAllenCarter_book}.
Fortunately, there is a way to circumvent the challenge.

One can calculate the bulk $F$ for small, finite and representative cells, and then partially correct for the concavity through a mean-field approach considering the interaction with identical nearby cells (self-interaction). The latter quantity can be seen as a cell viscosity that controls the transition from order to disorder. We call this the \underline{d}isorder \underline{v}iscosity \underline{c}orrection (\DVC) approach, based on finite cells. \DVC\ has two advantages: first, it prevents long-range fluctuation~\cite{kikuchi,Fontaine1979} leading to infinite phase separation~\cite{Kikuchi_JCP_1972,Cenedese_PHYSA_1994}; and second, the magnitude of the correction can be chosen to preserve the local concavity of the bulk $F$, which is necessary to allow the growth of surfaces/interfaces to perform their local stabilizing effect~\cite{curtarolo:art61,curtarolo:art36_etal}.

Despite this approximated approach cannot reproduce all the properties of a spinodal system, \DVC\ is still advantageous for discovering the existence and boundaries of spinodal regions inside the miscibility gap. Its immediate generalization to high-throughput and/or machine-learning searches~\cite{nmatHT,aflow4_etal} can expedite the discovery of microstructures in high-entropy materials.

\section{Disorder Viscosity Correction}
The \DVC\ workflow is introduced in Figure~\ref{fig:workflow}.
First, we define a macroscopic concentration grid $\{\mathbf{x}_1,\cdots,\mathbf{x}_{N_\sgrid}\}$, such that we uniformly sample the concentration space.
Then, at each $\mathbf{x}_i$, we fix the size of the micro-states using the \underline{p}artial \underline{o}ccupation formalism~\cite{curtarolo:art110} (\POCC\ tiles, \XXX) to the minimal possible dimension congruent with the concentration (Fig.~\ref{fig:workflow}(a)).
This choice corresponds to keeping only the shortest composition fluctuations possible --- resulting in the largest free energy hump possible and thus the upper bound of the spinodal transition temperature $T_\ssp$~\cite{Cenedese_PHYSA_1994}.
Next, at each $\mathbf{x}_i$, (Fig.~\ref{fig:workflow}(b)) from the ab initio energies of the \POCC\ tiles, (Fig.~\ref{fig:workflow}(c)) we compute the energy moments $M_\nu$ and cumulants $K_\nu$ which will be used to evaluate the free energy using the cumulant expansion.
In addition, at each $\mathbf{x}_i$, (Fig.~\ref{fig:workflow}(c)) we calculate the self-interaction energy of the disordered system $\Essi(T)$~\cite{Hoffman_MTR_1972}.
This quantity represents a self-coupling parameter of lattice dynamics (see Equations (11) and (24) in the original reference~\cite{Hoffman_MTR_1972}). Unfortunately, this quantity is extremely difficult to obtain numerically for every composition and temperature~\cite{Tsatskis_JPCM_1998}. Thus, we approximate it with the \POCC\ ensemble energy per atom at high-temperature $T_\sPOCC$, $\Essi(T)\approx\Essi\equiv\sum_iP_i(T_\sPOCC)E_i$, where $P_i$ and $E_i$ are the probability and energy of the $i$-tile, respectively (\XXX).
The approximation is exact in the $T\!\!\rightarrow\!\infty$ limit when tile-tile interactions can be neglected. Yet, both the chosen large $T_\sPOCC$ ($>$\SI{1000}{K}) and the lack of structural phase transition changing the lattice dynamics (solid$\leftrightarrow$liquid) make us confident that the level of approximation is appropriate for the purpose of our approach.

Our $\Essi$ has a more specific interpretation: its rate of change with respect to the size of the representative cells functions as a "viscous force" associated to the tendency of ordered cells becoming disordered; i.e., by forcing the nearby surrounding cell configurations to remain identical to the original cells.
At this point, (Fig.~\ref{fig:workflow}(d)) we have computed all the necessary quantities, $K_\nu$ and $\Essi$, at each of the concentration grid points, to fit these quantities using multivariate polynomials of the concentration $\mathbf{x}$ (\XXX).
This (Fig.~\ref{fig:workflow}(e)) gives analytical expressions for both the free energy $F(\mathbf{x},T)$ obtained using the cumulant expansion, and $\Essi(\mathbf{x})$ to easily compute $T_\ssp(\mathbf{x})$ from the locus of the determinant of $\partial^2 F/\partial x_i\partial x_j$ (concentrations' Hessian).
Likewise, using $T_\ssp(\mathbf{x})$, we calculate the maximum spinodal wavelength $\lambda_\ssp(\mathbf{x},T)$ using Cahn's approximation~\cite{Cahn1961795,Hoyt_ACTAMETALM_1990,curtarolo:art201_etal}.

Our choice of small \POCC\ tile sizes creates the upper bound of $T_\ssp$ given by the large concave free energy hump. As such, we correct $F$ by removing a fraction $\chi_\smix$ of $\Essi$ using the "viscosity" interpretation of $\Essi$, (Fig.~\ref{fig:workflow}(f)), similar in spirit to the self-interaction correction for electrons~\cite{Perdew_prb_1981} . We call this the disorder viscosity correction (\DVC).

The fraction $\chi_\smix$ is taken, as ansatz, (Fig.~\ref{fig:workflow}(g)) to be the average between the one giving maximum spinodal $T^\textrm{max}_\textrm{sp}$,
no correction $\Rightarrow$ $\chi(T^\textrm{max}_\textrm{sp})=0$, and the one giving complete miscibility $\Rightarrow$ $\chi(T^\textrm{min}_\textrm{sp}=0)$.
Thus, $\chi_\smix\equiv \chi(T^\textrm{min}_\textrm{sp}=0)/2$.
After determining $\chi_\smix$, (Fig.~\ref{fig:workflow}(h)) we compute the concentration-Hessian of the corrected free energy $F-\chi_\smix\Essi$ and (Fig.~\ref{fig:workflow}(i)) calculate the corrected $T_\ssp(\mathbf{x})$ and $\lambda_\ssp(\mathbf{x},T)$ .
Given the complexity of the whole process, the ultimate validity of the \DVC\ method is benchmarked against the available experimental measurements.

\begin{figure*}
    \includegraphics[width=0.95\linewidth]{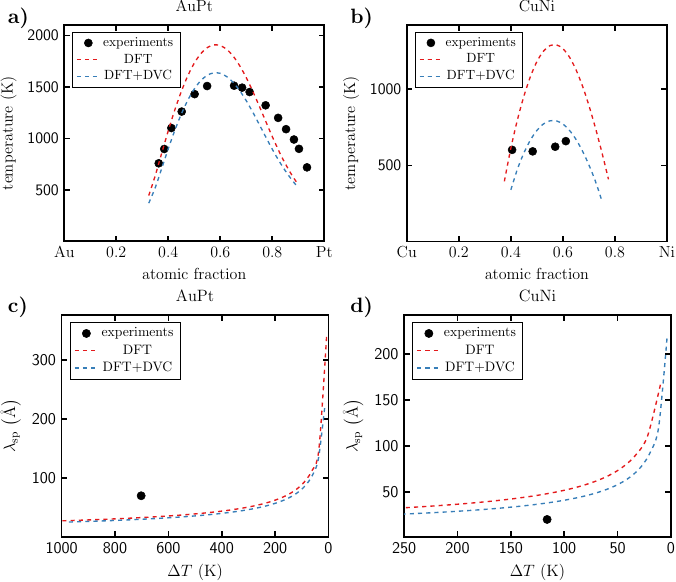}
    \caption{\small
    \textbf{Spinodal decomposition in binary systems.}
    (\textbf{a}–\textbf{b})
    Spinodal temperature, as a function of composition, for \ch{AuPt} and \ch{CuNi}.
    (\textbf{c}–\textbf{d})
    Maximum spinodal wavelength, as a function of the undercooling temperature, for 40/60 at.\% Au/Pt and 30/70 at.\% Cu/Ni alloys.
    The labels \DFT\ and \DFTDVC\ correspond to the results obtained without and with the \DVC\ to the free energy, respectively.
    \label{fig:binary}
    }
\end{figure*}

\noindent {\bf Calculating the free energy and spinodal temperature.}
Most thermodynamic quantities, \textit{in primis} $F$, can be calculated from the partition function $Z$~\cite{Landau_CTP5_SP_1969}:
\begin{equation}
    Z(\mathbf{x},T) \equiv e^{-F(\mathbf{x},T)/k_\mathrm{B}T}  \equiv \sum_{i}g_{i}e^{-E_{i}/k_\mathrm{B}T},
    \label{eq:part_func}
\end{equation}
where $\mathbf{x}$, $k_B$, $T$, $g_i$, and $E_i$ are the concentration vector, Boltzmann constant, temperature, factor-group degeneracy~\cite{curtarolo:art135_etal}, and energy of the $i$-configuration, respectively.
Generally, the summation in Eq.~(\ref{eq:part_func}) is quite challenging, especially for large systems, even using accelerated sampling approaches like in the Wang-Landau method~\cite{WangLandau_AJP2004}.
To address the issue, Kirkwood  proposed a moment expansion~\cite{Kirkwood_JCP_1938} (here written in modern notation and for an $n$-component disordered system):
\begin{equation}
    Z(\mathbf{x},T) = W\sum_{\nu=0}\frac{(-\beta)^\nu}{\nu!}M_\nu(\mathbf{x}),
    \label{eq:part_func_exp}
\end{equation}
with $\beta \equiv 1/k_\mathrm{B}T$.
Here, $W$ and $M_\nu$ are the total number of states and the $\nu$-th moment, respectively.
From combinatorial analysis, we have $W$:
\begin{equation*}
    W = \sum_ig_i = \frac{\left(\sum_k^nN_k\right)!}{\prod_k^n\left (N_k! \right)},
\end{equation*}
where $N_k$ is the total number of atoms of the $k$-species in the system. Obtaining the moments $M_\nu$ remains the culprit of the expansion.
For binary systems,
it is possible to derive an analytical expression for $M_\nu$ as a function of $\mathbf{x}$ and an ad-hoc interaction energy parameter~\cite{KrivoglazSmirnov64}.
However, it is still unclear if an analytical expression for ternaries and beyond can still be found. Even if it were, one would still need to evaluate the interaction energies between the different atomic species, a highly non-trivial task~\cite{Khachaturyan1978}.

To circumvent the impasse, we start from a Taylor expansion of Eq.~(\ref{eq:part_func}) as a function of $\beta$:
\begin{equation*}
    Z(\mathbf{x},T) = W\sum_{\nu=0}\frac{(-\beta)^\nu}{\nu!}\frac{\sum_ig_iE^\nu_i}{\sum_ig_i}.
    \label{eq:Ztaylor}
\end{equation*}
Here, the moments become $M_\nu= \sum_ig_iE^\nu_i/\sum_ig_i$ for any multi-component disordered system, and they do require the sum over \textit{all} the configurations.
Fortunately, averaged quantities, e.g.,
$M_\nu\equiv\langle\cdots\rangle$, tend to converge quickly, even with a coarse sampling of the phase space.
The latter is sampled through the \POCC\ formalism~\cite{curtarolo:art110} (\XXX).
Within \POCC, the moments become:
\begin{equation*}
    M^\sPOCC_\nu(\mathbf{x}) \approx \frac{\sum_i^{N_{\ssPOCC}}g_iE_i^\nu}{\sum_i^{N_\ssPOCC} g_i},
    \label{eq:mpocc}
\end{equation*}
where $N_\sPOCC$, $g_i$, and $E_i$ are the total number of \POCC\ tiles for the given concentration, the supercell degeneracy and the energy per atom of the $i$-tile, respectively~\cite{curtarolo:art135,curtarolo:art170}. Given the small size of the \POCC\ tiles, an ab initio parameterization becomes accessible~\cite{curtarolo:art110}.
{
Furthermore, the number of \POCC\ tiles required to evaluate the disordered system is typically one to two orders of magnitude smaller than that required for cluster expansion~\cite{clupan}.
}

\begin{figure}[ht!]
 \includegraphics[width=0.4\textwidth]{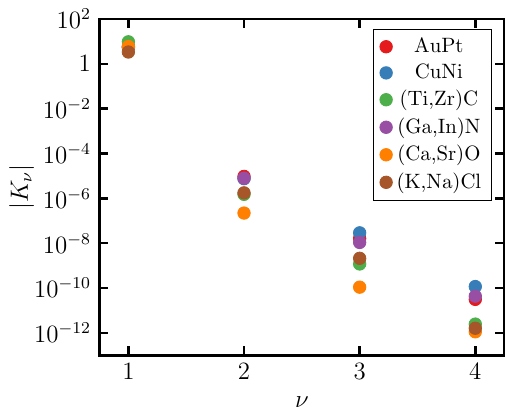}
 \vspace{-3mm}
 \caption{\small
    \textbf{Cumulant magnitude decay.} At equi-concentration, the magnitude of the cumulants decays as a function of the index for the disordered systems considered in this study. The large decrease in the magnitude is very general and observed at all concentrations.}
 \label{fig:cumulants_decay}
\end{figure}

The free energy can be evaluated through the logarithm of Eq.~(\ref{eq:part_func_exp}), leading to a cumulant expansion~\cite{Fisher_Cumulants}:
\begin{equation}
    -\beta F(\mathbf{x},T) = -\sum_k^n x_k \log (x_k) + \sum_{\nu=1}\frac{(-\beta)^\nu}{\nu!}K_\nu(\mathbf{x}),
    \label{eq:free_energy}
\end{equation}
where the terms are the Stirling approximation of $\log W$ and the $\nu$-cumulants ($K_\nu$).
The latter are known functions of the moments and are easy to calculate numerically~\cite{Rota_Cumulants}.
In addition, away from phase boundaries, cumulants decay rapidly in $\nu$, so one can consider only the first few terms~\cite{Kirkwood_JCP_1938,Chang_JCP_1941} (\XXX, Figure~\ref{fig:cumulants_decay}).
In operando, $F$ is evaluated at arbitrary $\mathbf{x}$ by fitting $n$-dimensional multivariate polynomials to the ab initio calculations performed on a concentration grid (\XXX).
{
We note that for multi-phase disordered systems, one needs to perform the POCC calculations for each parent lattice to obtain the correct thermodynamic density of state, and thus the free energy. Extensions to consider vibrational contributions to the POCC tiles can also be included by following the approach of Ref.~\onlinecite{curtarolo:art180}.
}

\begin{figure*}[ht!]
 \includegraphics[width=0.95\textwidth]{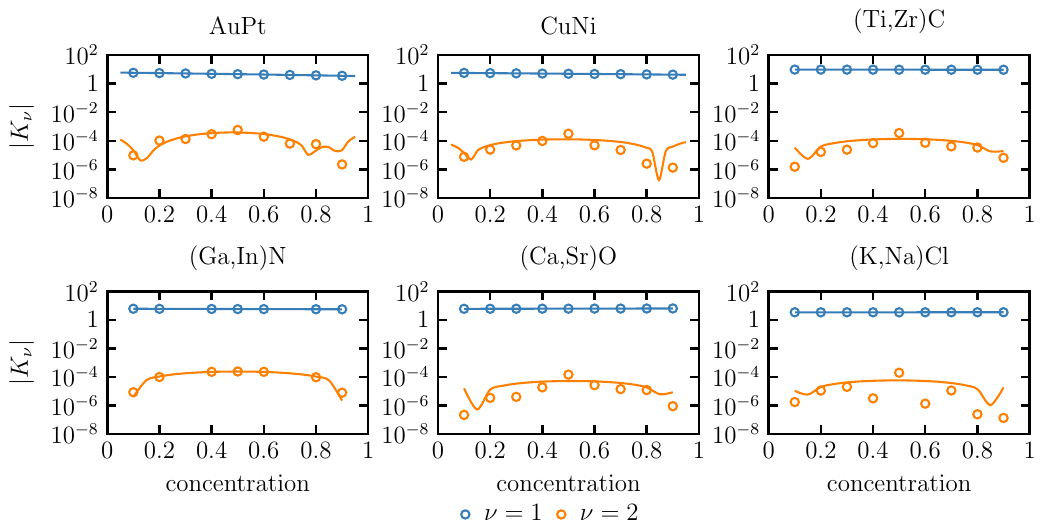}
  \vspace{-3mm}
 \caption{\small
    \textbf{Cumulant fitting.} Fitting of the cumulants using multivariate polynomials. The points are results obtained from the ab initio calculations, while the lines are the polynomial fits to the data.}
 \label{fig:cumulants_fit}
\end{figure*}

Then $T_\ssp$ is given by the locus of the determinant of the $(n-1)\times(n-1)$ stability matrix~\cite{DeFontaine_JCPSOL_1972}:
\begin{equation}
    \det \left [ D^2_{ij}F(\mathbf{x},T_\ssp) \right ] = 0,
    \label{eq:temp}
\end{equation}
where
\begin{equation*}
    D^2_{ij} = \frac{\partial^2}{\partial x_i \partial x_j} - \frac{\partial^2}{\partial x_i \partial x_n} - \frac{\partial^2 }{\partial x_n \partial x_j} + \frac{\partial^2}{\partial x_n \partial x_n}
\end{equation*}
is the concentration-Hessian defined by the fact that there are only $(n-1)$ independent concentrations.
The stability matrix is evaluated analytically through the fitting parameters of the multivariate polynomials (\XXX, Figure~\ref{fig:cumulants_fit}).
With these two ingredients, we solve Eq.~(\ref{eq:temp}) using Brent's method to find $T_\ssp$.
As stated previously, sampling the \POCC\ micro-states that are the minimal possible dimension congruent with the concentration yields the upper bound of the spinodal temperature $T^\textrm{max}_\textrm{sp}(\mathbf{x})$.

\begin{figure*}
    \includegraphics[width=0.95\linewidth]{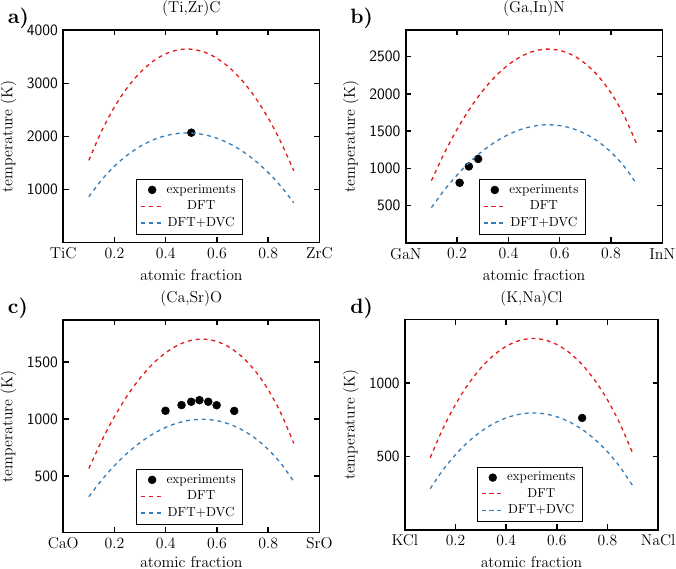}
    \caption{\small
    \textbf{Spinodal decomposition in ternary systems.} (\textbf{a}–\textbf{d}) Spinodal temperature, as a function of composition, for \ch{(Ti,Zr)C}, \ch{(Ga,In)N} \ch{(Ca,Sr)O}, and, \ch{(K,Na)Cl}.
    The labels \DFT\ and \DFTDVC\ correspond to the results obtained without and with the \DVC\ to the free energy, respectively.
    \label{fig:ternary}
    }
\end{figure*}
\noindent {\bf Calculating the self-consistent correction.}
We begin by computing the self-interaction energy $\Essi(\mathbf{x},T)$, which can be done in several ways~\cite{Brout_PPF_1967,Hoffman_MTR_1972,Fontaine1979}.
Within our computational schema, the most convenient method is by Hoffman~\cite{Hoffman_MTR_1972,DeFontaine_JCPSOL_1973}, where, in the infinite-temperature limit, $\Essi(\mathbf{x},T\!\!\rightarrow\!\infty) \equiv \Essi(\mathbf{x}) = \langle E \rangle(\mathbf{x})$, where $\langle E \rangle$ is the ensemble average energy of the disordered system.
This approximation works well away from any phase boundaries, an assumption that we used when we introduced the cumulant expansion for $F$.
{For example, the leading correction at order $\beta^{-1}$ is proportional to the variance of the effective cluster interactions (\ECI) in reciprocal space~\cite{Hoffman_MTR_1972}.
This correction is typically at least an of magnitude smaller than $\langle E \rangle$~\cite{atat1} (see Appendix).}
Furthermore, calculating $\Essi$ would have required volumetric integration of the pair-correlation functions of the disordered system; however, in the \POCC\ formalism, this quantity is evaluated directly, without any additional computational cost.
In operando, $\Essi$ is evaluated on the same concentration grid and fitted to a multivariate polynomial of the same degree as $K_1$ (cumulant with units of energy/atom, \XXX).

We have all the pieces for calculating the corrected $T_\ssp$ from the corrected free energy $F - \chi_\smix\Essi$, by resolving $\chi_\smix$ self-consistently.
Our procedure, outlined in Figure~\ref{fig:workflow}(f), begins by setting $\chi=0$ and computing $T_\ssp(\mathbf{x})$.
Next, if $T_\ssp(\mathbf{x})$ is finite on the concentration grid, then we increment $\chi$ by $\Delta \chi \ll 1$.
This is performed iteratively until $T_\ssp(\mathbf{x})=0$ on the whole concentration grid.
Then, we define $\chi_\smix$ as the midpoint between no \DVC\ ($\chi(T^\textrm{max}_\textrm{sp})=0$) and the value at which $T_\ssp$ vanishes ($\chi(T^\textrm{min}_\textrm{sp}=0)$).
The idea{, similar to the concept of the Hill average between the two limits of the elastic moduli~\cite{Hill_elastic_average_1952},} is that the former is an under-approximation of $\chi$, while the latter must be an over-approximation, since spinodal decomposition is not possible.
Thus, we take the average of the two bounds to obtain a pragmatic ansatz.
We use this value of $\chi_\smix\Essi$ as our \DVC\ when correcting the ab initio calculations.
{Further optimization of $\chi$ could be performed based on the properties of the disordered system, however we do not consider this in the current work.}

\section{Results and discussion}

\noindent {\bf Comparison to experiments: binaries.}
We validate our methodology using two systems, \ch{AuPt} and \ch{CuNi}, which were selected due to the availability of extensive experimental data on the spinodal temperature and measurements of the spinodal wavelength for specific concentrations~\cite{Okamoto_BAPD_1985,Vrijen_PRB_1978,Singhal1978_JACRYST_1978,AvilaDavila_JAC_2008}.
In addition, both of these systems have asymmetrical spinodal curves; i.e., must be modeled by sub-regular solutions.
The comparison of $T_\ssp$, between the theory and experiment, shown in Figure~\ref{fig:binary}(a-b), represents good agreement on the position of the spinodal curve in concentration space, even with the uncorrected $\Delta F$.
Once we include the correction to $\Delta F$ by $\chi_\smix\Essi$, we recover the appropriate transition temperature for the alloys.
{The under-approximation of $T_\text{sp}$ for \ch{CuNi} far from the center of the spinodal curve is attributed to the extraction of the experimental value from the Clapp–Moss model, which is known to systematically overestimate the transition temperature~\cite{deFontaine_SCRMET_1986}.}
In fact, other recent experiments on the miscibility gap show that the curve does not {extend over such a large concentration range and is much narrower~\cite{Iguchi_ACTAMAT_2018}}, in strong agreement with our findings.

For the prediction of the spinodal wavelength $\lambda_\ssp$ (\XXX), shown in Figure~\ref{fig:binary}(c-d), we see that we can only capture a rough estimate of the wavelength and \DVC\ does not have much influence on the results.
The origin of this discrepancy is usually attributed to the regular solid-solution approximation~\cite{Lass_CALPHAD_2006}, but it is much more subtle.
In the regular solid-solution approximation, the gradient energy coefficient matrix $\boldsymbol\kappa$ can be directly calculated from the interatomic potential, which gives a rather accurate estimate for $\lambda_\ssp$~\cite{curtarolo:art201}.
Alternatively, if more accuracy is required, it can be calculated from lengthy molecular dynamics~\cite{Hoyt_PRB_2007} or Monte Carlo simulations~\cite{Asta_ACTAMAT_2000}.
If the information about the interatomic potential is unavailable, then $\boldsymbol\kappa$ can be deduced from the regular solid-solution energy parameters~\cite{cahn_hilliard_1,Hildebrand_JCP_1933}, as it was done in \XXX.
This is the part of the assumption that leads to the primary deviation in $\lambda_\ssp$ from the correct value.
Nevertheless, since robust interatomic potentials for disordered systems are not available in the overwhelming majority of cases, our method to calculate $\lambda_\ssp$ can be used as a first step to screen materials of interest.

\noindent {\bf Comparison to experiments: ternaries.}
Thus far we have presented a comparison between theory and experiment only for binary alloys.
While our approach uses \DFT, which is readily accessible, it can be argued that other computational methods~\cite{kikuchi,Mohri_MMTRA_2017,Korringa1947392} can already achieve similar results.
However, we note that those methods are limited to systems that have muffin-tin type potentials~\cite{Papanikolaou_JPCM_2002} and/or where the interaction between the lattice sites is short-ranged~\cite{An_JSP_1988} --- largely metallic alloys.
On the other hand, our approach is applicable to any {disordered material (solid-solutions) where formation energies of “lattice-occupation” configurations can be reliably calculated or approximated.
Within the context of \DFT, one should be careful when dealing with system that contain: long-range dispersion forces, delocalization errors and strong correlations~\cite{Cohen_CHEMREV_2011}.
Since the method involves the second derivative of energy with respect to the concentration, any error in the \DFT\ energy that is non-linear with the concentration will influence the results presented (non-linear corrections, such as the coordination-corrected energy method~\cite{curtarolo:art150} can be trained and employed as needed).
}
To test this claim, we calculate the spinodal decomposition in a diverse set of ternary disordered systems where experimental measurements were present.

We first consider the ceramic \ch{(Ti,Zr)C}, a cubic system, but with significant directionality in the bonding~\cite{Lowther_JOPACOS_1987}.
While the experimental data is extremely scarce~\cite{Ma_IJRMHM_2016}, we find excellent agreement between the available experiments and our theory, shown in Figure~\ref{fig:ternary}(a), once the correction is included.
In addition, the spinodal curve is symmetric, in correspondence with the nearly-symmetric miscibility curve, obtained from first-principles thermodynamic calculations~\cite{Borgh_ACTAMAT_2014}.

Second, we consider the ceramic \ch{(Ga,In)N}, also with a strong directional bonding, and a macroscopic polarization due to its non-cubic wurtzite crystal structure~\cite{Sahoo_JAP_2013}.
Again, we find very good agreement between our calculations and the experimental measurements~\cite{Stringfellow_JJCG_2016}, shown in Figure~\ref{fig:ternary}(b).
Furthermore, where experiments were absent, our spinodal curve agrees well with the one obtained from an empirical valence force field model~\cite{Takayama_JAP_2001}.

Next, we consider the metal oxide \ch{(Ca,Sr)O}, a system with ionic bonding where long range electrostatic forces are present.
Additionally, ab initio energies are systematically under-approximated, attributed to the over-binding of the \ch{O2} molecule~\cite{Jones_RMP_1989}.
While beyond-\DFT\ methods exist to correct this error~\cite{Yan_formation_PRB_2013}, they are computationally expensive and are not suitable for high-throughput approaches~\cite{nmatHT}.
Yet, given this impediment, we find that our results are only a modest under-approximation of the experimental values~\cite{Jacob_JACERS_1998}, shown in Figure~\ref{fig:ternary}(c).

Finally, we consider \ch{(K,Na)Cl}, an alkali halide salt, which is also an ionic solid.
Likewise, such a system presents a challenge to ab initio due to the strong van der Waals interaction between the ion cores~\cite{Tang_MRE_2018}.
Given the very limited experimental data, we again find only a slight under-approximation in our results, shown in Figure~\ref{fig:ternary}(d).
More accurate results could probably be achieved using dispersion-corrected ab initio, which improves the description of the van der Waals interactions~\cite{Antony_PCCP_2006}.

Overall we find great agreement for the ternary disordered systems between the calculated spinodal curves and the experimental measurements found in the literature.
We note that, within our method, further improvements to this prediction can be made by optimizing $\chi_\smix$.
Regrettably, we have not found any bulk measurements of the spinodal wavelength for these materials and thus we cannot make any claims about our predictions in these systems.
Nevertheless, our results for the binary systems indicate that the spinodal wavelength calculations should provide a good estimate.
In general, detailed spinodal decomposition experiments for disordered ternary systems have been very sparse in the literature.
However, given the diverse examples presented, we have highlighted the versatility and accuracy of our approach compared to existing methods.

\section{\XXX}
\noindent{\bf \POCC\ method.} Within the \POCC\ formalism, a disordered system is modeled as an ensemble average of $N_\sPOCC$ tiles, orthogonally extracted from a Hermite supercell-volume, with probabilities given by~\cite{curtarolo:art110}:
\begin{equation*}
    P_i(T_\sPOCC) = \frac{g_ie^{-E_i/k_\mathrm{B}T_\sPOCC}}{\sum_{i}^{N_\ssPOCC}g_ie^{-E_i/k_\mathrm{B}T}},
\end{equation*}
where $g_i$ and $E_i$ are the supercell degeneracy and energy of the $i$-tile, respectively~\cite{curtarolo:art135,curtarolo:art170}.
Here, $k_\mathrm{B}$ is the Boltzmann constant and $T_\sPOCC$ is a virtual temperature describing how much disorder has been statistically explored~\cite{curtarolo:art200_etal}.
An observable of the disordered system $\langle O \rangle$ is given by an ensemble average over the tile quantities $O_i$, such that:
\begin{equation*}
    \langle O \rangle(T_\sPOCC) = \sum_{i}^{N_\ssPOCC}P_i(T_\sPOCC)O_i.
\end{equation*}

\noindent{\bf Multivariate polynomial fitting.} Consider a multivariate function of $n$ variables $f(x_1,x_2,\cdots,x_{n-1},x_n)\equiv f(\mathbf{x})$ that is evaluated on a grid of $N_\sgrid$ points $\{\mathbf{x}_1, \mathbf{x}_2,\cdots,\mathbf{x}_{N_\sgrid-1},\mathbf{x}_{N_\sgrid}\}$, which we write as:
\begin{align*}
    \mathbf{f} &=
    \begin{bmatrix}
        f(\mathbf{x}_1) \\
        f(\mathbf{x}_2) \\
        \vdots \\
        f(\mathbf{x}_{N_\sgrid-1}) \\
        f(\mathbf{x}_{N_\sgrid})
    \end{bmatrix},
\end{align*}
which is a $(N_\sgrid \times 1)$ column vector.
To fit these values to a $n$-dimensional multivariate polynomial of degree $d$:
\begin{equation*}
    \mathcal{P}_d(\mathbf{x})=\sum_{i_1=0}^{d}\cdots\sum_{i_n=0}^{d}\omega_{i_1, \cdots, i_n}x_1^{i_1}\cdots x_n^{i_n},
\end{equation*}
where $\{\omega\}$ are the fitting parameters.
Let us define the following:
\begin{align*}
    \boldsymbol\omega &=
    \begin{bmatrix}
        \omega_{0,0, \cdots, 0,0} \\
        \omega_{1,0, \cdots, 0,0}  \\
        \vdots \\
        \omega_{d,d, \cdots, d,d-1}  \\
        \omega_{d,d, \cdots, d,d}
    \end{bmatrix},
\end{align*}
which is a $((d+1)^n \times 1)$ column vector, and the multivariate Vandermonde matrix:
\begin{widetext}
\begin{align*}
    \mathbf{X} &=
    \begin{bmatrix}
        \prod\mathbf{x}_1^{\odot [0,0,\cdots,0,0]} & \prod\mathbf{x}_1^{\odot [1,0,\cdots,0,0]} & \cdots & \prod\mathbf{x}_1^{\odot [d,d,\cdots,d,d-1]} & \prod\mathbf{x}_1^{\odot [d,d,\cdots,d,d]} \\
        \prod\mathbf{x}_2^{\odot [0,0,\cdots,0,0]} & \prod\mathbf{x}_2^{\odot [1,0,\cdots,0,0]} & \cdots & \prod\mathbf{x}_2^{\odot [d,d,\cdots,d,d-1]} & \prod\mathbf{x}_2^{\odot [d,d,\cdots,d,d]} \\
        \vdots & \vdots & \ddots & \vdots & \vdots\\
        \prod\mathbf{x}_{N_g-1}^{\odot [0,0,\cdots,0,0]} & \prod\mathbf{x}_{N_g-1}^{\odot [1,0,\cdots,0,0]} & \cdots & \prod\mathbf{x}_{N_g-1}^{\odot [d,d,\cdots,d,d-1]} & \prod\mathbf{x}_{N_g-1}^{\odot [d,d,\cdots,d,d]} \\
        \prod\mathbf{x}_{N_g}^{\odot [0,0,\cdots,0,0]} & \prod\mathbf{x}_{N_g}^{\odot [1,0,\cdots,0,0]} & \cdots & \prod\mathbf{x}_{N_g}^{\odot [d,d,\cdots,d,d-1]} & \prod\mathbf{x}_{N_g}^{\odot [d,d,\cdots,d,d]}
    \end{bmatrix},
\end{align*}
\end{widetext}
which is a $(N_g \times (d+1)^n)$ matrix.
The symbol $\odot$ refers to an element-wise or Hadamard power, such that:
\begin{equation*}
    \prod\mathbf{a}^{\odot\mathbf{b}} \equiv \prod_ia_i^{b_i},
\end{equation*}
for vectors $\mathbf{a}$ and $\mathbf{b}$.
Then, the fitting parameters are given by the following unique solution:
\begin{equation*}
    \boldsymbol\omega = (\mathbf{X}^T\mathbf{X})^{-1}\mathbf{X}^T\mathbf{f}.
\end{equation*}

For the cumulants $K_\nu$, the order of the polynomial $d^{(\nu)}$ is chosen to be the largest value, such that, the concentration-Hessian of the spinodal temperature $T_\ssp(\mathbf{x})$ is negative semi-definite, for the uncorrected free energy ($\chi=0$).
From our experience, we have found that $d^{(\nu)} \le 2$ produces physical spinodal curves, whereas higher polynomial orders yield spurious oscillations (Runge's phenomenon) in the spinodal temperature.

\noindent{\bf Maximum spinodal wavelength.}
The equation for the maximum spinodal wavelength has been derived in multiple references~\cite{Cahn1961795,Hoyt_ACTAMETALM_1990,curtarolo:art201}.
Written in our notation, for the multi-component case:
\begin{equation*}
    \lambda_\ssp(\mathbf{x}, T) = \sqrt{16\pi^2\max_{ij}\left ( -\frac{\kappa_{ij}(\mathbf{x}, T)}{D^2_{ij}F(\mathbf{x},T)}\right )},
    \label{eq:lambda}
\end{equation*}
where $\boldsymbol\kappa$ is the gradient energy coefficient matrix.
Since we have already calculated the denominator, the only difficulty in evaluating this expression is computing $\boldsymbol\kappa$.
For a cubic lattice, $\kappa_{ij}=\kappa_{ii}\delta_{ij}$ (for non-cubic lattices, see Ref.~\citenum{Hillert_ACTAMETAL_1977}) and in the regular solid-solution approximation:
\begin{equation*}
    \kappa_{ii}(\mathbf{x}, T) = \frac{1}{2}\Omega_{in}(T)\xi(\mathbf{x})^2,
    \label{eq:lambda_regsol}
\end{equation*}
where $\Omega_{in}$ and $\xi$ are the regular solid-solution energy parameter between species $i$ and $n$, and the effective interaction distance, respectively.
Taking advantage of the polynomial expansion of $K_\nu$ and $\mathcal{E}_\ssi$, the energy parameters are given explicitly by:
\begin{equation*}
    \Omega_{in}(T) = \sum_{\nu=1}\frac{(-\beta)^{\nu - 1}}{\nu!}\tilde{\omega}^{(\nu)}_{in},
\end{equation*}
where $\{\tilde{\omega}^{(\nu)}\}$ are the multivariate polynomial fitting parameters for the regular solid-solution.

Meanwhile, the effective interaction distance is approximated by $\xi(\mathbf{x}) = (11/7)^{1/2}\langle r\rangle(\mathbf{x})$, where $\langle r\rangle$ is the ensemble averaged nearest neighbor distance of the disordered system~\cite{cahn_hilliard_1,cahn_hilliard_2}.
Calculating $\langle r\rangle$ within the \POCC\ formalism is not straightforward because the \POCC\ tiles do not necessarily share the same symmetry as the parent lattice.
One would either need to use crystal-structure specific formulas for $r$, or transform the lattice vectors~\cite{gus_enum} into more convenient ones where the nearest neighbor distance formula is known, for every single \POCC\ tile --- both of which are very cumbersome.
Instead, we calculate $\langle r\rangle$ through another intensive quantity:
$$
    \langle\rho\rangle(\mathbf{x}) = {N_\mathrm{cell}} /{V_\mathrm{cell}(\langle r\rangle)},
$$
where $\langle\rho\rangle$, $N_\mathrm{cell}$, $V_\mathrm{cell}$ are the ensemble averaged number density of the disordered system, number of atoms in the unit cell of the parent lattice and unit cell volume of the parent lattice, respectively.
It is trivial to calculate $\langle\rho\rangle$ for any crystal structure. For cubic parent lattices, the values $\{N_\mathrm{cell},V_\mathrm{cell}\}$ are well-known to be $\{1,\langle r\rangle^3\}$, $\{2,8\langle r\rangle^3/\sqrt{27}\}$, and $\{4,\sqrt{8}\langle r\rangle^3\}$ for simple, body-centered and face-centered cubic, respectively.

\noindent{\bf Computational details.} All ab initio calculations were performed using \VASP~\cite{vasp} under the \AFLOW\ standard~\cite{curtarolo:art104_etal}.
We used the Perdew-Burke-Ernzerhof functional (\PBE)~\cite{PBE}, except for \ch{(Ca,Sr)O} where we used the variant for solids~\cite{Perdew_PRL_2008_etal} due to its ability to more accurately model the oxides.

\begin{figure*}[ht!]
 \includegraphics[width=0.95\textwidth]{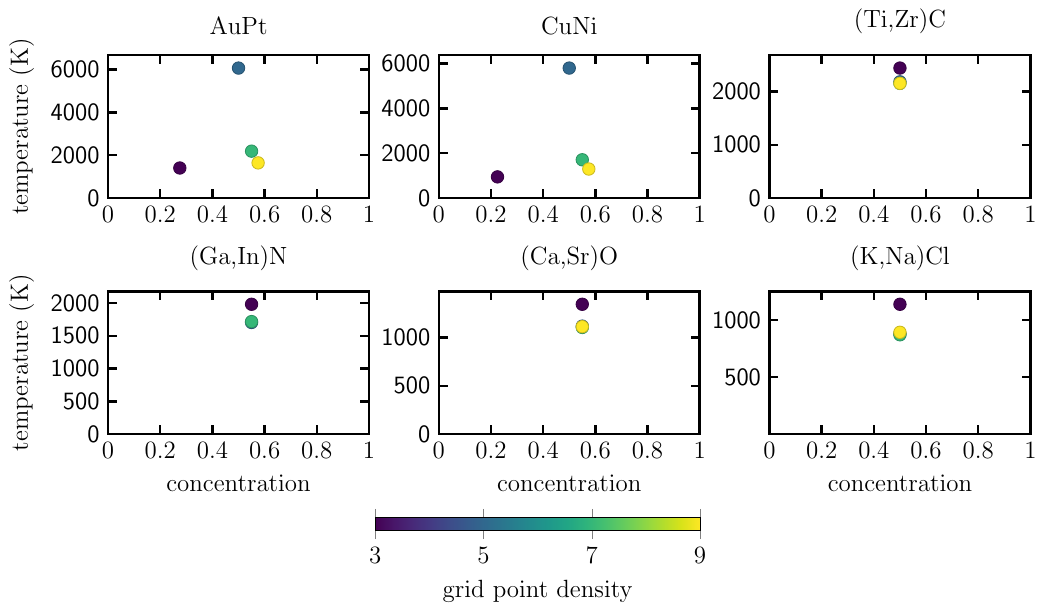}
 \vspace{-3mm}
 \caption{\small
    \textbf{Concentration grid convergence.} The convergence of the maximum of $T_\ssp$ in $(\mathbf{x},T)$ space as a function of the grid density.}
 \label{fig:grid_conv}
\end{figure*}

The concentration grid, where the ab initio energies where evaluated, for the $cF4$ $AB$ systems, was \{[0.1, 0.9], [0.2, 0.8], [0.3, 0.7], [0.4, 0.6], [0.5, 0.5], [0.6, 0.4], [0.7, 0.3], [0.8, 0.2], [0.9, 0.1]\}.
Meanwhile, for the $cF8$ $ABC$ systems, where $C$ is the anion, we evaluated energies on \{[0.05, 0.45, 0.5], [0.1, 0.4, 0.5], [0.15, 0.35, 0.5], [0.2, 0.3, 0.5], [0.25, 0.25, 0.5], [0.3, 0.2, 0.5], [0.35, 0.15, 0.5], [0.4, 0.1, 0.5], [0.05, 0.45, 0.5]\}.
While, for the $hP4$ $ABC$ system, we omitted the points \{[0.15, 0.35, 0.5], [0.35, 0.15, 0.5]\} due to the computational cost of requiring over \num{10000} ab initio calculations each.
{
The convergence of the maximum of $T_\ssp$ in $(\mathbf{x},T)$ space as a function of the grid density is shown in Fig.~\ref{fig:grid_conv}.
We observe convergence with a grid density of $\geq7$ for all systems.
}

\section{Conclusions}
In this work we have introduced the disorder viscosity correction (\DVC) as a simplified yet powerful approach to computationally model spinodal decomposition.
\DVC\ corrects the locally concave bulk free energy derived from finite, small representative cells, by a fraction of the self-interaction energy of the solid solution.
This energy represents the cost to constrict disorder into smaller cells.
By preventing long-range concentration fluctuations and correcting the local bulk concavity, \DVC\ allows interfacial/curvature-driven free energy stabilization, enabling spinodal modeling without the computational overhead of full ab initio treatments.
Our approach offers a practical and scalable pathway for high-throughput and machine-learning-driven materials discovery, particularly in the multi-component landscape of compositionally complex and high-entropy materials.

{
\section{Appendix}
Similar to the cumulant expansion of the free energy, the self-interaction energy can be expanded in a cumulant expansion (c.f. Eq.~25 in Ref.~\cite{Hoffman_MTR_1972}):
\begin{equation}
    \Essi(\mathbf{x},T) = \langle E \rangle(\mathbf{x}) + \sum_{\nu=1}a_\nu(\mathbf{x})(-\beta)^\nu \mathcal{K}_\nu(\mathbf{x}),
    \label{eq:e_si}
\end{equation}
where $a_\nu$, $\beta \equiv 1/k_\mathrm{B}T$ and $\mathcal{K}_\nu$ are the $\nu$-th coefficient, inverse temperature and the $\nu$-th cumulant, respectively.
Note that $a_\nu$ are only functions of the product of $\mathbf{x}$, so they are at most $\mathcal{O}(1)$.
We will only consider the first term in the sum because it is the largest in magnitude~\cite{Kirkwood_JCP_1938}.
From Eqs.~(14)b and (27)a in Ref.~\cite{Hoffman_MTR_1972}, we see that $\mathcal{K}_1$ is equal to the variance of the reciprocal space \ECI.
Usually, the \ECI\ are calculated and reported in real space, so we will derive a relationship between the variance of the real space \ECI\ and the reciprocal one.
Furthermore, we will only consider \ECI\ pairs, as was done in Ref.~\cite{Hoffman_MTR_1972}, to keep the results analytical.
Then, the reciprocal space \ECI\ $\tilde{J}$ is defined as a sum on the shells at radius $r$:
\begin{equation*}
    \tilde{J}(\mathbf{k}) = \sum_{|\mathbf{r}|=r}J_r\cos (\mathbf{k}\cdot\mathbf{r}),
\end{equation*}
where $J_r$ is the real space \ECI\ (defined only on the shells).
Note that within the context of the real space \ECI, $\langle E \rangle = -\sum_{r\neq 0}J_r$~\cite{Hoffman_MTR_1972}.
Next, exploiting the properties of the cosine, we write:
\begin{align*}
    \langle \tilde{J}(\mathbf{k})\rangle_\mathbf{k} &= J_0\\
    \langle \tilde{J}(\mathbf{k})^2\rangle_\mathbf{k} &= \sum_rw_rJ_r^2,
\end{align*}
where $w_r$ is the weight of the shell at $r$.
Finally, we observe that the variance of the \ECI\ in reciprocal space $\mathcal{K}_1$ is equal to the variance of the \ECI\ in real space:
\begin{equation*}
    \langle \tilde{J}(\mathbf{k})^2\rangle_\mathbf{k}-\langle \tilde{J}(\mathbf{k})\rangle_\mathbf{k}^2 = \sum_rw_rJ_r^2 - J_0^2 \equiv \langle J_r^2 \rangle_r - \langle J_r \rangle_r^2.
\end{equation*}
Note that this relationship does not hold if we include triplets and higher clusters because the sum used to define $\tilde{J}(\mathbf{k})$ will include more complicated trigonometric functions than just cosines -- although the variance of the real space \ECI\ can still be computed numerically from the reciprocal space \ECIs.
Nevertheless, the variance of the real space \ECI\ would not change orders of magnitude if triplets and higher clusters were included (upper bound given by the Rayleigh-Ritz theorem~\cite{Gould95}).
}

{
Using the values for the real space \ECI\ pairs listed in Ref.~\cite{atat1} for a diverse set of binary systems (\ch{GeSi}, \ch{(Ca,Mg)O}, \ch{AlTi}, and \ch{AuCu}), we calculate the magnitude of $\langle E \rangle$ and $\mathcal{K}_1$ as shown in Table~\ref{tab:si_compare}.
The last column in Table~\ref{tab:si_compare}, is the first term in the sum of Eq.~\ref{eq:e_si} evaluated at equi-composition and room temperature.
We observe a relative difference of at least an order of magnitude between $\langle E \rangle$ and that term.
}

\begin{table}[h]
\centering
\begin{tabular}{llll}
\hline
\\[-2.0ex]
System & $|\langle E \rangle|$ & $|\mathcal{K}_1|$ & $a_1|\mathcal{K}_1|/k_\mathrm{B}T_\mathrm{room}$ \\
\\[-2.0ex]
\hline
\\[-2.5ex]
\ch{GeSi} & \num{3.6} & \num{0.4} & \num{6.3e-4} \\
\\[-2.5ex]
\ch{(Ca,Mg)O} & \num{10.9} & \num{233.7} & \num{0.3} \\
\\[-2.5ex]
\ch{AlTi} (hcp) & \num{109.6} & \num{418.0} & \num{0.5}  \\
\\[-2.5ex]
\ch{AlTi} (fcc) & \num{39.7} & \num{1326.4} & \num{1.7} \\
\\[-2.5ex]
\ch{AuCu} & \num{0.5} & \num{40.5} & \num{5.1e-2} \\
\\[-2.5ex]
\hline
\end{tabular}
\caption{{Calculation of the terms in the self-interaction energy using the pair effective cluster interaction (\ECI) energies obtained from Ref.~\cite{atat1}. The coefficient $a_1$, evaluated at equi-composition, is equal to $1/2^5=0.03125$ and the thermal energy at room temperature ($k_\mathrm{B}T_\mathrm{room}$) is \SI{25}{\milli\electronvolt\per atom}. The units for $\langle E \rangle$ and $\mathcal{K}_1$ are \SI{}{\milli\electronvolt\per atom} and \SI{}{(\milli\electronvolt\per atom)^2}, respectively.}
}
\label{tab:si_compare}
\end{table}

\subsection*{CRediT authorship contribution statement}
{\small
\textbf{Simon Divilov:} Writing, methodology, investigation, analysis, conceptualization.
\textbf{Hagen Eckert:} Analysis.
\textbf{Nico Hotz, Xiomara Campilongo:} Review \& editing.
\textbf{Stefano Curtarolo:} Writing, methodology, review \& editing, supervision, conceptualization, project administration.
}
\subsection*{Declaration of competing interest}
{\small
The authors declare that they have no competing financial interests
or personal relationships that could have appeared to influence the work
reported in this article.
The author Dr. Stefano Curtarolo is an Editor for Acta Materialia and was not involved in the editorial review or the decision to publish this article.
}
\subsection*{Acknowledgments}
{\small
The authors thank Arrigo Calzolari, Ohad Levy and Amir Natan for fruitful discussions.
This research was supported by the Office of Naval Research under grant N00014-21-1-2515 and N00014-23-1-2615.
This work received high-performance computer time and resources from the DoD High Performance Computing Modernization Program (Frontier).
We acknowledge Auro Scientific, LLC for computational support.
}
\subsection*{Data availability}
{\small
The data that support the findings in this project are available from the corresponding author upon
request.
}

\newcommand{\Ozolins}{Ozoli{\c{n}}{\v{s}}}


\begin{thebibliography}{10}
\expandafter\ifx\csname urlstyle\endcsname\relax
  \providecommand{\doi}[1]{doi:\discretionary{}{}{}#1}\else
  \providecommand{\doi}{doi:\discretionary{}{}{}\begingroup
  \urlstyle{rm}\Url}\fi
\providecommand{\selectlanguage}[1]{\relax}
\providecommand{\bibAnnoteFile}[1]{%
  \IfFileExists{#1}{\begin{quotation}\noindent\textsc{Key:} #1\\
  \textsc{Annotation:}\ \input{#1}\end{quotation}}{}}
\providecommand{\bibAnnote}[2]{%
  \begin{quotation}\noindent\textsc{Key:} #1\\
  \textsc{Annotation:}\ #2\end{quotation}}

\bibitem{Cahn1961795}
J.~W. Cahn, \emph{On spinodal decomposition}, Acta\ Metall. \textbf{9},
  795--801 (1961).
\input{Cahn1961795}

\bibitem{curtarolo:art80_etal}
{G.~S. Rohrer \textit{et al.}}, \emph{Challenges in Ceramic Science: A Report
  from the Workshop on Emerging Research Areas in Ceramic Science}, J.\ Am.\
  Ceram.\ Soc. \textbf{95}, 3699--3712 (2012).
\input{curtarolo:art80_etal}

\bibitem{Cahn_ActaMetall_SpinodalHardening_1963}
J.~W. Cahn, \emph{Hardening by spinodal decomposition}, Acta\ Metall.
  \textbf{11}, 1275--1282 (1963).
\input{Cahn_ActaMetall_SpinodalHardening_1963}

\bibitem{Findik_MATTDES_2012}
F.~Findik, \emph{Improvements in spinodal alloys from past to present}, Mater.\
  Des. \textbf{42}, 131--146 (2012).
\input{Findik_MATTDES_2012}

\bibitem{Dietl_RMP_2015_etal}
{T. Dietl {\it et al.}}, \emph{Spinodal nanodecomposition in semiconductors
  doped with transition metals}, Rev.\ Mod.\ Phys. \textbf{87}, 1311--1377
  (2015).
\input{Dietl_RMP_2015_etal}

\bibitem{Marro_PRB_1975}
J.~Marro, A.~B. Bortz, M.~H. Kalos, and J.~L. Lebowitz, \emph{Time evolution of
  a quenched binary alloy. {II.} {C}omputer simulation of a three-dimensional
  model system}, Phys.\ Rev.\ B \textbf{12}, 2000--2011 (1975).
\input{Marro_PRB_1975}

\bibitem{Langer_PRA_1975}
J.~S. Langer, M.~Bar-on, and H.~D. Miller, \emph{New computational method in
  the theory of spinodal decomposition}, Phys.\ Rev.\ A \textbf{11}, 1417--1429
  (1975).
\input{Langer_PRA_1975}

\bibitem{Xiong_SSP_2011}
W.~Xiong, K.~A. Gr\"{o}nhagen, J.~{\AA}gren, M.~Selleby, J.~Odqvist, and
  Q.~Chen, \emph{Investigation of spinodal decomposition in Fe-Cr alloys:
  {CALPHAD} modeling and phase field simulation}, Solid State Phenomena
  \textbf{172--174}, 1060--1065 (2011).
\input{Xiong_SSP_2011}

\bibitem{Zhou_CALPHAD_2017}
J.~Zhou, J.~Zhong, L.~Chen, L.~Zhang, Y.~Du, Z.-K. Liu, and P.~H. Mayrhofer,
  \emph{Phase equilibria, thermodynamics and microstructure simulation of
  metastable spinodal decomposition in c--Ti$_{1-x}$Al$_x$N coatings}, Calphad
  \textbf{56}, 92--101 (2017).
\input{Zhou_CALPHAD_2017}

\bibitem{Deffrennes_MATTDES_2022}
G.~Deffrennes, K.~Terayama, T.~Abe, and R.~Tamura, \emph{A machine
  learning--based classification approach for phase diagram prediction},
  Mater.\ Des. \textbf{215}, 110497 (2022).
\input{Deffrennes_MATTDES_2022}

\bibitem{curtarolo:art201_etal}
{S. Divilov {\it et al.}}, \emph{{A priori procedure to establish spinodal
  decomposition in alloys}}, Acta Mater. \textbf{266}, 119667 (2024).
\input{curtarolo:art201_etal}

\bibitem{deFontaine_ssp_1994}
D.~{de Fontaine}, \emph{Cluster Approach to Order-Disorder Transformations in
  Alloys}, in \emph{Solid State Physics}, edited by H.~Ehrenreich and
  D.~Turnbull (Academic Press, New York, 1994), vol.~47, pp. 33--176,
  \doi{10.1016/S0081-1947(08)60639-6}.
\input{deFontaine_ssp_1994}

\bibitem{kikuchi}
R.~Kikuchi, \emph{A Theory of Cooperative Phenomena}, Phys.\ Rev. \textbf{81},
  988 (1951).
\input{kikuchi}

\bibitem{nmatHT}
S.~Curtarolo, G.~L.~W. Hart, M.~{Buongiorno Nardelli}, N.~Mingo, S.~Sanvito,
  and O.~Levy, \emph{The high-throughput highway to computational materials
  design}, Nat.\ Mater. \textbf{12}, 191--201 (2013).
\input{nmatHT}

\bibitem{curtarolo:art114_etal}
{P. Nath {\it et al.}}, \emph{High-throughput prediction of finite-temperature
  properties using the quasi-harmonic approximation}, Comput.\ Mater.\ Sci.
  \textbf{125}, 82--91 (2016).
\input{curtarolo:art114_etal}

\bibitem{clupan}
A.~Seko, Y.~Koyama, and I.~Tanaka, \emph{Cluster expansion method for
  multicomponent systems based on optimal selection of structures for
  density-functional theory calculations}, Phys.\ Rev.\ B \textbf{80}, 165122
  (2009).
\input{clupan}

\bibitem{deFontaine_Clustering_1975}
D.~{de Fontaine}, \emph{Clustering Effects in Solid Solutions} (Springer US,
  1975), p. 129–178, \doi{10.1007/978-1-4757-1120-2_3}.
\input{deFontaine_Clustering_1975}

\bibitem{Korringa1947392}
J.~Korringa, \emph{On the calculation of the energy of a {B}loch wave in a
  metal}, Physica \textbf{13}, 392--400 (1947).
\input{Korringa1947392}

\bibitem{Landau_CTP5_SP_1969}
L.~D. Landau and E.~M. Lifshitz, \emph{Statistical Physics}, in \emph{Course of
  Theoretical Physics} (Pergamon Press, 1969), vol.~5.
\input{Landau_CTP5_SP_1969}

\bibitem{Kinetic_of_Materials}
R.~W. Balluffi, S.~M. Allen, and W.~G. Carter, eds., \emph{Kinetic of
  Materials} (Wiley, 2005).
\input{Kinetic_of_Materials}

\bibitem{Langer_AnnPhys_1971}
J.~S. Langer, \emph{Theory of spinodal decomposition in alloys}, Annals of
  Physics \textbf{65}, 53--86 (1971).
\input{Langer_AnnPhys_1971}

\bibitem{cahn_hilliard_1}
J.~W. Cahn and J.~E. Hilliard, \emph{Free Energy of a Nonuniform System. I.
  Interfacial Free Energy}, J.\ Chem.\ Phys. \textbf{28}, 258--267 (1958).
\input{cahn_hilliard_1}

\bibitem{lupis}
C.~H.~P. Lupis, \emph{Chemical Thermodynamics of Materials} (North-Holland, New
  York, 1983).
\input{lupis}

\bibitem{Kikuchi_JCP_1972}
R.~Kikuchi, \emph{Boundary Free Energy in the Lattice Model. {III}. Solution of
  the Paradox}, J.\ Chem.\ Phys. \textbf{57}, 787--791 (1972).
\input{Kikuchi_JCP_1972}

\bibitem{Cenedese_PHYSA_1994}
P.~Cenedese and R.~Kikuchi, \emph{Numerical limit of the spinodal point},
  Physica A: Statistical Mechanics and its Applications \textbf{205}, 747--755
  (1994).
\input{Cenedese_PHYSA_1994}

\bibitem{Kikuchi_JCP_1967}
R.~Kikuchi, \emph{Cooperative Phenomena in the Triangular Lattice}, J.\ Chem.\
  Phys. \textbf{47}, 1664--1668 (1967).
\input{Kikuchi_JCP_1967}

\bibitem{BaluffiAllenCarter_book}
R.~W. Balluffi, S.~M. Allen, and W.~C. Carter, \emph{Kinetics of Materials}
  (John Wiley \& Sons, Inc., Hoboken, New Jersey, 2005),
  \doi{10.1002/0471749311}.
\input{BaluffiAllenCarter_book}

\bibitem{Fontaine1979}
D.~D. Fontaine, \emph{Configurational Thermodynamics of Solid Solutions},
  \emph{Solid State Physics}, vol.~34 (Academic Press, 1979),
  \doi{10.1016/S0081-1947(08)60360-4}.
\input{Fontaine1979}

\bibitem{curtarolo:art61}
R.~V. Chepulskii and S.~Curtarolo, \emph{{\it Ab Initio} Insights on the Shapes
  of Platinum Nanocatalysts}, ACS\ Nano \textbf{5}, 247--254 (2011).
\input{curtarolo:art61}

\bibitem{curtarolo:art36_etal}
{A.~R. Harutyunyan {\it et al.}}, \emph{Reduced Carbon Solubility in {Fe}
  Nanoclusters and Implications for the Growth of Single-Walled Carbon
  Nanotubes}, Phys.\ Rev.\ Lett. \textbf{100}, 195502 (2008).
\input{curtarolo:art36_etal}

\bibitem{aflow4_etal}
{S. Divilov {\it et al.}}, \emph{AFLOW4: Heading Toward Disorder}, High Entropy
  Alloys Mater. \textbf{3}, 178--187 (2025).
\input{aflow4_etal}

\bibitem{curtarolo:art110}
K.~Yang, C.~Oses, and S.~Curtarolo, \emph{Modeling Off-Stoichiometry Materials
  with a High-Throughput \textit{Ab-Initio} Approach}, Chem.\ Mater.
  \textbf{28}, 6484--6492 (2016).
\input{curtarolo:art110}

\bibitem{Hoffman_MTR_1972}
D.~W. Hoffman, \emph{Configurational entropy and solute correlation in
  disordered alloys}, Metall.\ Trans. \textbf{3}, 3231--3238 (1972).
\input{Hoffman_MTR_1972}

\bibitem{Tsatskis_JPCM_1998}
I.~Tsatskis, \emph{Theory of the temperature dependence of the
  Fermi-surface-induced splitting of the alloy diffuse-scattering intensity
  peak}, J.\ Phys.:\ Condens.\ Matter \textbf{10}, L145--L151 (1998).
\input{Tsatskis_JPCM_1998}

\bibitem{Hoyt_ACTAMETALM_1990}
J.~J. Hoyt, \emph{The continuum theory of nucleation in multicomponent
  systems}, Acta Metallurgica et Materialia \textbf{38}, 1405--1412 (1990).
\input{Hoyt_ACTAMETALM_1990}

\bibitem{Perdew_prb_1981}
J.~P. Perdew and A.~Zunger, \emph{Self-interaction correction to
  density-functional approximations for many-electron systems}, Phys.\ Rev.\ B
  \textbf{23}, 5048--5079 (1981).
\input{Perdew_prb_1981}

\bibitem{curtarolo:art135_etal}
{D. Hicks {\it et al.}}, \emph{\textit{AFLOW-SYM}: platform for the complete,
  automatic and self-consistent symmetry analysis of crystals}, Acta\
  Crystallogr.\ Sect.\ A \textbf{74}, 184--203 (2018).
\input{curtarolo:art135_etal}

\bibitem{WangLandau_AJP2004}
D.~P. Landau, S.-H. Tsai, and M.~Exler, \emph{{A new approach to Monte Carlo
  simulations in statistical physics: Wang-Landau sampling}}, Am.\ J.\ Phys.
  \textbf{72}, 1294--1302 (2004).
\input{WangLandau_AJP2004}

\bibitem{Kirkwood_JCP_1938}
J.~G. Kirkwood, \emph{Order and Disorder in Binary Solid Solutions}, J.\ Chem.\
  Phys. \textbf{6}, 70--75 (1938).
\input{Kirkwood_JCP_1938}

\bibitem{KrivoglazSmirnov64}
M.~A. Krivoglaz and A.~A. Smirnov, \emph{The Theory of Order-Disorder in
  Alloys} (Macdonald, Materials Park, London, 1964).
\input{KrivoglazSmirnov64}

\bibitem{Khachaturyan1978}
A.~G. Khachaturyan, \emph{Ordering in substitutional and interstitial solid
  solutions}, Prog.\ Mater.\ Sci. \textbf{22}, 1--150 (1978).
\input{Khachaturyan1978}

\bibitem{curtarolo:art135}
D.~Hicks, C.~Oses, E.~Gossett, G.~Gomez, R.~H. Taylor, C.~Toher, M.~J. Mehl,
  O.~Levy, and S.~Curtarolo, \emph{\textit{AFLOW-SYM}: platform for the
  complete, automatic and self-consistent symmetry analysis of crystals}, Acta\
  Crystallogr.\ Sect.\ A \textbf{74}, 184--203 (2018).
\input{curtarolo:art135}

\bibitem{curtarolo:art170}
D.~Hicks, C.~Toher, D.~C. Ford, F.~Rose, C.~{De Santo}, O.~Levy, M.~J. Mehl,
  and S.~Curtarolo, \emph{{AFLOW-XtalFinder}: a reliable choice to identify
  crystalline prototypes}, npj\ Comput.\ Mater. \textbf{7}, 30 (2021).
\input{curtarolo:art170}

\bibitem{Fisher_Cumulants}
R.~A. Fisher and J.~Wishart, \emph{The Derivation of the Pattern Formulae of
  Two-Way Partitions from those of Simpler Patterns}, Proceedings of the London
  Mathematical Society \textbf{s2-33}, 15--208 (1932).
\input{Fisher_Cumulants}

\bibitem{Rota_Cumulants}
G.-C. Rota and J.~Shen, \emph{On the Combinatorics of Cumulants}, Journal of
  Combinatorial Theory, Series A \textbf{91}, 283--304 (2000).
\input{Rota_Cumulants}

\bibitem{Chang_JCP_1941}
T.~S. Change, \emph{A Note on Bethe-Kirkwood's Partition Function for a Binary
  Solid Solution}, J.\ Chem.\ Phys. \textbf{9}, 169--174 (1941).
\input{Chang_JCP_1941}

\bibitem{curtarolo:art180}
M.~Esters, C.~Oses, D.~Hicks, M.~J. Mehl, M.~Jahn{\'{a}}tek, M.~D. Hossain,
  J.-P. Maria, D.~W. Brenner, C.~Toher, and S.~Curtarolo, \emph{Settling the
  matter of the role of vibrations in the stability of high-entropy carbides},
  Nat.\ Commun. \textbf{12}, 5747 (2021).
\input{curtarolo:art180}

\bibitem{DeFontaine_JCPSOL_1972}
D.~{de Fontaine}, \emph{An analysis of clustering and ordering in
  multicomponent solid solutions-{I.} Stability criteria}, J.\ Phys.\ Chem.\
  Solids \textbf{33}, 297--310 (1972).
\input{DeFontaine_JCPSOL_1972}

\bibitem{Brout_PPF_1967}
R.~Brout and H.~Thomas, \emph{Molecular field theory, the Onsager reaction
  field and the spherical model}, Physics Physique Fizika \textbf{3}, 317--329
  (1967).
\input{Brout_PPF_1967}

\bibitem{DeFontaine_JCPSOL_1973}
D.~{de Fontaine}, \emph{An analysis of clustering and ordering in
  multicomponent solid solutions-{II.} Stability criteria}, J.\ Phys.\ Chem.\
  Solids \textbf{34}, 1285--1304 (1973).
\input{DeFontaine_JCPSOL_1973}

\bibitem{atat1}
A.~{{v}an {d}e Walle} and G.~Ceder, \emph{Automating First-Principles Phase
  Diagram Calculations}, J.\ Phase Equilib. \textbf{23}, 348--359 (2002).
\input{atat1}

\bibitem{Hill_elastic_average_1952}
R.~Hill, \emph{The elastic behaviour of a crystalline aggregate}, Proc. Phys.
  Soc. Sect. A \textbf{65}, 349 (1952).
\input{Hill_elastic_average_1952}

\bibitem{Okamoto_BAPD_1985}
H.~Okamoto and T.~B. Massalski, \emph{The {Au}-{Pt} (Gold-Platinum) system},
  Bull.\ Alloy\ Phase\ Diag. \textbf{6}, 46--56 (1985).
\input{Okamoto_BAPD_1985}

\bibitem{Vrijen_PRB_1978}
J.~Vrijen and S.~Radelaar, \emph{Clustering in {Cu}-{Ni} alloys: A diffuse
  neutron-scattering study}, Phys.\ Rev.\ B \textbf{17}, 409--421 (1978).
\input{Vrijen_PRB_1978}

\bibitem{Singhal1978_JACRYST_1978}
S.~P. Singhal, H.~Herman, and G.~Kostorz, \emph{Neutron small-angle scattering
  study of phase decomposition in {Au}{\textendash}{Pt}}, J.\ Appl.\
  Crystallogr. \textbf{11}, 572--577 (1978).
\input{Singhal1978_JACRYST_1978}

\bibitem{AvilaDavila_JAC_2008}
E.~O. Avila-Davila, V.~Lopez-Hirata, M.~L. Saucedo-Mu{\~{n}}oz, and J.~L.
  Gonzalez-Velazquez, \emph{Microstructural simulation of phase decomposition
  in {Cu}{\textendash}{Ni} alloys}, J.\ Alloys\ Compd. \textbf{460}, 206--212
  (2008).
\input{AvilaDavila_JAC_2008}

\bibitem{deFontaine_SCRMET_1986}
D.~de~Fontaine, A.~Finel, and T.~Mohri, \emph{Comparison of short-range order
  intensity obtained by Monte Carlo simulation and by the cluster variation
  method}, Scr.\ Metall. \textbf{20}, 1045--1047 (1986).
\input{deFontaine_SCRMET_1986}

\bibitem{Iguchi_ACTAMAT_2018}
Y.~Iguchi, G.~L. Katona, C.~Cserháti, G.~A. Langer, and Z.~Erdélyi, \emph{On
  the miscibility gap of {Cu-Ni} system}, Acta\ Mater. \textbf{148}, 49--54
  (2018).
\input{Iguchi_ACTAMAT_2018}

\bibitem{Lass_CALPHAD_2006}
E.~A. Lass, W.~C. Johnson, and G.~J. Shiflet, \emph{Correlation between
  {CALPHAD} data and the Cahn–Hilliard gradient energy coefficient and
  exploration into its composition dependence}, Calphad \textbf{30}, 42--52
  (2006).
\input{Lass_CALPHAD_2006}

\bibitem{curtarolo:art201}
S.~Divilov, H.~Eckert, C.~Toher, R.~Friedrich, A.~C. Zettel, D.~W. Brenner,
  W.~G. Fahrenholtz, D.~E. Wolfe, E.~Zurek, J.-P. Maria, N.~Hotz,
  X.~Campilongo, and S.~Curtarolo, \emph{{A priori procedure to establish
  spinodal decomposition in alloys}}, Acta Mater. \textbf{266}, 119667 (2024).
\input{curtarolo:art201}

\bibitem{Hoyt_PRB_2007}
J.~J. Hoyt, \emph{Molecular dynamics study of equilibrium concentration
  profiles and the gradient energy coefficient in {Cu-Pb} nanodroplets}, Phys.\
  Rev.\ B \textbf{76}, 094102 (2007).
\input{Hoyt_PRB_2007}

\bibitem{Asta_ACTAMAT_2000}
M.~Asta and J.~J. Hoyt, \emph{Thermodynamic properties of coherent interfaces
  in f.c.c.-based {Ag–Al} alloys: a first-principles study}, Acta\ Mater.
  \textbf{48}, 1089--1096 (2000).
\input{Asta_ACTAMAT_2000}

\bibitem{Hildebrand_JCP_1933}
J.~H. Hildebrand and S.~E. Wood, \emph{The Derivation of Equations for Regular
  Solutions}, J.\ Chem.\ Phys. \textbf{1}, 817--822 (1933).
\input{Hildebrand_JCP_1933}

\bibitem{Mohri_MMTRA_2017}
T.~Mohri, \emph{Cluster Variation Method as a Theoretical Tool for the Study of
  Phase Transformation}, Metall.\ Mater.\ Trans.\ A \textbf{48}, 2753--2770
  (2017).
\input{Mohri_MMTRA_2017}

\bibitem{Papanikolaou_JPCM_2002}
N.~Papanikolaou, R.~Zeller, and P.~H. Dederichs, \emph{Conceptual improvements
  of the {KKR} method}, J.\ Phys.:\ Condens.\ Matter \textbf{14}, 2799--2823
  (2002).
\input{Papanikolaou_JPCM_2002}

\bibitem{An_JSP_1988}
G.~An, \emph{A note on the cluster variation method}, Journal of Statistical
  Physics \textbf{52}, 727--734 (1988).
\input{An_JSP_1988}

\bibitem{Cohen_CHEMREV_2011}
A.~J. Cohen, P.~Mori-S\'{a}nchez, and W.~Yang, \emph{Challenges for density
  functional theory}, Chem.\ Rev. \textbf{112}, 289--320 (2011).
\input{Cohen_CHEMREV_2011}

\bibitem{curtarolo:art150}
R.~Friedrich, D.~Usanmaz, C.~Oses, A.~Supka, M.~Fornari, M.~{Buongiorno
  Nardelli}, C.~Toher, and S.~Curtarolo, \emph{Coordination corrected ab initio
  formation enthalpies}, npj\ Comput.\ Mater. \textbf{5}, 59 (2019).
\input{curtarolo:art150}

\bibitem{Lowther_JOPACOS_1987}
J.~E. Lowther and A.~Andriotis, \emph{Directionality of the metallic bonding in
  titanium carbide}, J.\ Phys.\ Chem.\ Solids \textbf{48}, 713--717 (1987).
\input{Lowther_JOPACOS_1987}

\bibitem{Ma_IJRMHM_2016}
T.~Ma, R.~Borrajo-Pelaez, P.~Hedstr\"{o}m, I.~Borgh, A.~Blomqvist, S.~Norgren,
  and J.~Odqvist, \emph{Microstructure evolution during phase separation in
  {Ti}-{Zr}-{C}}, Int.\ J.\ Refract.\ Metals\ Hard\ Mater. \textbf{61},
  238--248 (2016).
\input{Ma_IJRMHM_2016}

\bibitem{Borgh_ACTAMAT_2014}
I.~Borgh, P.~Hedstr\"{o}m, A.~Blomqvist, J.~Ågren, and J.~Odqvist,
  \emph{Synthesis and phase separation of {(Ti, Zr)C}}, Acta\ Mater.
  \textbf{66}, 209--218 (2014).
\input{Borgh_ACTAMAT_2014}

\bibitem{Sahoo_JAP_2013}
S.~K. Sahoo, B.~K. Sahoo, and S.~Sahoo, \emph{The macroscopic polarization
  effect on thermal conductivity of binary nitrides}, J.\ Appl.\ Phys.
  \textbf{114}, 163501 (2013).
\input{Sahoo_JAP_2013}

\bibitem{Stringfellow_JJCG_2016}
G.~B. Stringfellow, \emph{Thermodynamics of {III-V} and {III-Nitride} alloys},
  Japanese Journal of Crystal Growth \textbf{43}, 222--232 (2016).
\input{Stringfellow_JJCG_2016}

\bibitem{Takayama_JAP_2001}
T.~Takayama, M.~Yuri, K.~Itoh, and J.~S. Harris, \emph{Theoretical predictions
  of unstable two-phase regions in wurtzite group-{III}-nitride-based ternary
  and quaternary material systems using modified valence force field model},
  J.\ Appl.\ Phys. \textbf{90}, 2358--2369 (2001).
\input{Takayama_JAP_2001}

\bibitem{Jones_RMP_1989}
R.~O. Jones and O.~Gunnarsson, \emph{The density functional formalism, its
  applications and prospects}, Rev.\ Mod.\ Phys. \textbf{61}, 689--746 (1989).
\input{Jones_RMP_1989}

\bibitem{Yan_formation_PRB_2013}
J.~Yan, J.~S. Hummelsh{\o}j, and J.~K. N{\o}rskov, \emph{Formation energies of
  group {I} and {II} metal oxides using random phase approximation}, Phys.\
  Rev.\ B \textbf{87}, 075207 (2013).
\input{Yan_formation_PRB_2013}

\bibitem{Jacob_JACERS_1998}
K.~T. Jacob and Y.~Waseda, \emph{Solid‐State Immiscibility and Thermodynamics
  of the Calcium Oxide‐Strontium Oxide System}, J.\ Am.\ Ceram.\ Soc.
  \textbf{81}, 1065--1068 (1998).
\input{Jacob_JACERS_1998}

\bibitem{Tang_MRE_2018}
H.~Tang and J.~Tao, \emph{Comparative study of the properties of ionic solids
  from density functionals}, Materials Research Express \textbf{5}, 076302
  (2018).
\input{Tang_MRE_2018}

\bibitem{Antony_PCCP_2006}
J.~Antony and S.~Grimme, \emph{Density functional theory including dispersion
  corrections for intermolecular interactions in a large benchmark set of
  biologically relevant molecules}, Phys.\ Chem.\ Chem.\ Phys. \textbf{8}, 5287
  (2006).
\input{Antony_PCCP_2006}

\bibitem{curtarolo:art200_etal}
{S. Divilov {\it et al.}}, \emph{{Disordered enthalpy-entropy descriptor for
  high-entropy ceramics discovery}}, Nature \textbf{625}, 66--73 (2024).
\input{curtarolo:art200_etal}

\bibitem{Hillert_ACTAMETAL_1977}
M.~Hillert and M.~Jarl, \emph{A regular-solution model for interstitial
  solutions in {HCP} metals}, Acta\ Metall. \textbf{25}, 1--9 (1977).
\input{Hillert_ACTAMETAL_1977}

\bibitem{cahn_hilliard_2}
J.~W. Cahn and J.~E. Hilliard, \emph{Free Energy of a Nonuniform System. III.
  Nucleation in a Two-Component Incompressible Fluid}, J.\ Chem.\ Phys.
  \textbf{31}, 688--699 (1959).
\input{cahn_hilliard_2}

\bibitem{gus_enum}
G.~L.~W. Hart and R.~W. Forcade, \emph{Algorithm for generating derivative
  structures}, Phys.\ Rev.\ B \textbf{77}, 224115 (2008).
\input{gus_enum}

\bibitem{vasp}
G.~Kresse and J.~Furthm{\"{u}}ller, \emph{Efficient iterative schemes for {\it
  ab initio} total-energy calculations using a plane-wave basis set}, Phys.\
  Rev.\ B \textbf{54}, 11169--11186 (1996).
\input{vasp}

\bibitem{curtarolo:art104_etal}
{C.~E. Calderon {\it at al.}}, \emph{The {AFLOW} standard for high-throughput
  materials science calculations}, Comput.\ Mater.\ Sci. \textbf{108}, 233--238
  (2015).
\input{curtarolo:art104_etal}

\bibitem{PBE}
J.~P. Perdew, K.~Burke, and M.~Ernzerhof, \emph{Generalized Gradient
  Approximation Made Simple}, Phys.\ Rev.\ Lett. \textbf{77}, 3865--3868
  (1996).
\input{PBE}

\bibitem{Perdew_PRL_2008_etal}
{J.~P. Perdew {\it et al.}}, \emph{Restoring the Density-Gradient Expansion for
  Exchange in Solids and Surfaces}, Phys.\ Rev.\ Lett. \textbf{100}, 136406
  (2008).
\input{Perdew_PRL_2008_etal}

\bibitem{Gould95}
S.~H. Gould, \emph{Variational methods for eigenvalue problems} (Dover
  Publications, Mineola, NY, 1995).
\input{Gould95}

\end{thebibliography}
\end{document}